\documentclass[journal,twoside,web]{ieeecolor}
\usepackage{tmi}
\usepackage{cite}
\usepackage{amsmath,amssymb,amsfonts}
\usepackage{mathtools}
\usepackage{algorithmic}
\usepackage{graphicx}
\usepackage{textcomp}
\usepackage{tabularx}
\usepackage{bm}
\usepackage{makecell}
\usepackage{array}
\usepackage{float}
\usepackage{ulem}
\usepackage{subcaption} 
\usepackage{caption}
\usepackage{xcolor}
\usepackage[
colorlinks,
linkcolor=blue,
anchorcolor=blue,
filecolor=blue,
urlcolor=blue,
citecolor=blue]{hyperref}
\usepackage[capitalise]{cleveref}
\Crefname{section}{Sec.}{Secs.}
\Crefname{table}{Tab.}{Tabs.}

\renewcommand{\textcolor}[2]{#2}
\def\BibTeX{{\rm B\kern-.05em{\sc i\kern-.025em b}\kern-.08em
    T\kern-.1667em\lower.7ex\hbox{E}\kern-.125emX}}
\begin{document}
\title{3D MedDiffusion: A 3D Medical Latent Diffusion Model for Controllable and High-quality Medical Image Generation}

\author{{
Haoshen Wang,
Zhentao Liu,
Kaicong Sun,
Xiaodong Wang,
Dinggang Shen, \IEEEmembership{Fellow, IEEE},
Zhiming Cui
}
\thanks{This work was supported in part by NSFC under Grant 6230012077, 
and in part by Shanghai Municipal Central Guided Local Science and Technology Development Fund under Project YDZX20233100001001. 
It was also supported by the High Performance Computing (HPC) Platform of ShanghaiTech University.\textit{(Corresponding author: Zhiming Cui.)}}
\thanks{Haoshen Wang, Zhentao Liu, Kaicong Sun, Dinggang Shen and Zhiming Cui are with the School of Biomedical Engineering \& State Key Laboratory of Advanced Medical Materials and Devices, ShanghaiTech Univerisity, Shanghai, 201210, China. Dinggang Shen is also with Shanghai United Imaging Intelligence Co., Ltd., Shanghai, 200230, China, and Shanghai Clinical Research and Trial Center, Shanghai, 201210, China. (e-mail: \{wanghsh2022, liuzht2022, sunkc, dgshen, cuizhm\}@shanghaitech.edu.cn). }
\thanks{Xiaodong Wang is with United Imaging Healthcare Co., Ltd., Shanghai, 201807,
	China  (e-mail: xiaodong.wang@united-imaging.com).}
}

\maketitle

\begin{abstract}
The generation of medical images presents significant challenges due to their high-resolution and three-dimensional nature. Existing methods often yield suboptimal performance in generating high-quality 3D medical images, and there is currently no universal generative framework for medical imaging.
In this paper, we introduce a 3D Medical Latent Diffusion (3D MedDiffusion) model for controllable, high-quality 3D medical image generation. 
3D MedDiffusion incorporates a novel, highly efficient Patch-Volume Autoencoder that compresses medical images into latent space through patch-wise encoding and recovers back into image space through volume-wise decoding.
Additionally, we design a new noise estimator to capture both local details and global structural information during diffusion denoising process.
3D MedDiffusion can generate fine-detailed, high-resolution images (up to 512x512x512) and effectively adapt to various downstream tasks as it is trained on large-scale datasets covering CT and MRI modalities and different anatomical regions (from head to leg).
Experimental results demonstrate that 3D MedDiffusion surpasses state-of-the-art methods in generative quality and exhibits strong generalizability across tasks such as sparse-view CT reconstruction, fast MRI reconstruction, and data augmentation for segmentationand classification. 
Source code and checkpoints are available at \href{https://github.com/ShanghaiTech-IMPACT/3D-MedDiffusion}{https://github.com/ShanghaiTech-IMPACT/3D-MedDiffusion}.

\end{abstract}


% \vspace{-5mm}
\section{Introduction}
\label{sec:introduction}
\IEEEPARstart{G}{enerative} artificial intelligence (AI) has led to remarkable breakthroughs in recent years, driving substantial advancements across various domains and applications. 
Specifically, it has significantly contributed to progress in medical imaging and healthcare, with immense potential for future innovations. 
Generative AI have shown strong capabilities in various medical applications including image translation~\cite{ozbey2023unsupervised}, segmentation~\cite{amit2021segdiff}, and the reconstruction of Computed Tomography (CT)~\cite{liu2023dolce} and Magnetic Resonance (MR) images~\cite{cao2024high}.

Different kinds of generative models have shaped such rapid advancement including Variational Autoencoders (VAE)~\cite{kingma2013auto}, normalizing flows~\cite{rezende2015variational}, Generative Adversarial Networks (GAN)~\cite{goodfellow2020generative}, and Denoising Diffusion Probabilistic Model (DDPM)~\cite{ho2020denoising}.
VAEs and normalizing flows are likelihood-based models, which explicitly model the data distribution and are theoretically well-founded. However, they often produce relatively modest generative results. 
On the other hand, GANs are generative models that employ an adversarial training approach to learn the underlying probability distribution of the data.
This method can result in training instability, characterized by issues such as mode collapse~\cite{che2016mode}. 
DDPM addresses these limitations by utilizing a diffusion process, modeled as a Markov chain~\cite{sohl2015deep,ho2020denoising}.
Noise is incrementally added to the data, and a model is trained to reverse this process to generate samples. 
This method ensures stable training, producing high-quality, diverse outputs while offering a interpretable generative process.
Large-scale models like stable diffusion~\cite{rombach2022high} are widely applied in natural image domains and easily adapted for tasks such as image inpainting~\cite{saharia2022palette}, and super-resolution~\cite{wang2024exploiting}.

Despite recent advancements, several challenges continue to impede the broader application of generative models in medical imaging. 
First, adapting computer vision algorithms to medical imaging is challenging due to its three-dimensional nature, which introduces significant computational overhead.
Existing generative models in the medical domain are often limited in both resolution and quality~\cite{friedrich2024wdm,khader2022medical}.
Some works rely on slice-wise 2D generation~\cite{zhu2023make,volokitin2020modelling,han2023medgen3d}, while clinical practice needs 3D volumetric imaging.
Second, the lack of efficient and controllable generation mechanisms forces the development of task-specific models, thereby increasing development costs.

We propose 3D Medical Latent Diffusion (3D MedDiffusion), a large-scale 3D generative model designed for high-fidelity generation of medical images. 
First, 3D MedDiffusion introduces a novel Patch-Volume Autoencoder that compresses high-resolution images into low-resolution latent representations via a patch-wise encoder and recovers the entire image with a volume-wise decoder.
% Patch-wise encoding reduces the computational overhead of volume-wise encoding by exploiting the local properties in 3D medical images, while volume-wise decoding ensures artifact-free reconstruction results.
Patch-wise encoding reduces the computational overhead in the training stage by exploiting the local properties in 3D medical images, while volume-wise decoding ensures artifact-free reconstruction results.
Second, we design a novel noise estimator (named BiFlowNet) to replace the standard U-Net~\cite{ronneberger2015u} in diffusion model. 
This module combines intra-patch and inter-patch flows to process both local (patch-wise) and global (volume-wise) information. 
The dual flow structure is crucial for reconstructing local details from noise while preserving global structure information.
Furthermore, 3D MedDiffusion is trained on a comprehensive and diverse collection of 3D medical image datasets, covering both CT and MRI modalities and multiple anatomical regions from head to leg. 
Notably, ControlNet~\cite{zhang2023adding}, a widely used method for injecting conditions into diffusion models, can be seamlessly integrated with the pre-trained 3D MedDiffusion, facilitating efficient fine-tuning for various downstream tasks.

In this study, we evaluate 3D MedDiffusion on its generative performance and its effectiveness in downstream tasks: sparse-view CT reconstruction, fast MRI reconstruction, and data augmentation for segmentation and classification. 
3D MedDiffusion outperforms existing methods, demonstrating its superior performance.

\section{Related work}
\subsection{Medical Image Generation}
Medical image generation is promising to address several clinical tasks, such as modality translation, CT enhancement, and MR acceleration.
Existing research has explored the application of GAN-based methods for generating 3D medical images. 
Jin et al.~\cite{jin2019applying} introduced an autoencoding GAN to generate 3D brain MRI images. 
Cirillo et al.~\cite{cirillo2021vox2vox} proposed a 3D model conditioned on multi-channel 3D brain MR images to generate tumor masks for segmentation. 
Sun et al.~\cite{sun2022hierarchical} developed a GAN with a hierarchical design to address memory demands of generating high-resolution 3D images.

Nevertheless, the unstable training process inherent to GAN framework often results in mode collapse and low quality generation. 
Recently, many researchers have shifted their focus to diffusion models (DMs) due to their high fidelity generation and stable training~\cite{ho2020denoising}. 
Müller-Franzes et al.~\cite{muller2023multimodal} proposed a conditional latent DM, demonstrating its promise for generating 2D medical images. 
Moghadam et al.~\cite{moghadam2023morphology} showed that DMs were capable of generating a wide range of histopathological images.
However, 3D imaging is more prevalent in medical contexts.
This task poses challenges due to high computational complexity of DMs for three-dimensional data.
Consequently, many researchers explored efficient methods for generating 3D medical images using DMs. 
Zhu et al.~\cite{zhu2023make} developed 3D image generators using pseudo-3D architectures, addressing volumetric inconsistencies through refiners or tuning. 
Friedrich et al.~\cite{friedrich2024wdm} applied a diffusion model to wavelet-decomposed images for efficiency. 
Khader et al.~\cite{khader2022medical} employed a latent DM combining an autoencoder and video DM to generate high-quality 3D CT and MR images.
Guo et al.~\cite{guo2024maisi} introduced MAISI, a concurrent latent DM based framework that improved autoencoder efficiency via tensor-splitting parallelism.
This method divides feature maps into smaller segments, enabling parallel processing across multiple GPUs or sequential processing on a single GPU.
It resembles sliding window inference on feature space, so it produces less artifacts than directly applying it on image space.
In contrast, our Patch-Volume Autoencoder follows a two-stage patch-wise and volume-wise training strategy, ensuring no boundary artifacts during single-GPU inference. 
Besides, we can also integrate tensor parallelism~\cite{shoeybi2019megatron} for multi-GPU inference to accelerate the process.
\vspace{-5.3pt}
\subsection{Medical Generative Model for Downstream Tasks}
Medical generative models hold great promise for diverse downstream tasks~\cite{friedrich2024deep,fan2019adversarial,xiang2018deep}.
Recent studies have explored the data augmentation with DMs, such as training conditional DMs to generate synthetic data from segmentation masks~\cite{guo2024maisi}. 
Other works have applied DMs to solve inverse problems, including CT reconstruction~\cite{chung2023decomposed} and image-to-image translation~\cite{kim2024adaptive}.
However, most of these methods require dedicated training for specific tasks. 
In contrast, our method leverages ControlNet~\cite{zhang2023adding} to eliminate task-specific training, enabling efficient fine-tuning for diverse downstream tasks.

%%Method...........
\section{Method}
Our approach consists of three key components: a Patch-Volume Autoencoder that compresses a 3D medical image into a compact latent space (\cref{sec:vqgan}), a noise estimator architecture named BiFlowNet for the 3D medical diffusion model (\cref{sec:dm}), and ControlNet for conditional generation in downstream tasks (\cref{sec:controlnet}).

\begin{figure*}[!t]
\centerline{\includegraphics[width=0.98\textwidth]{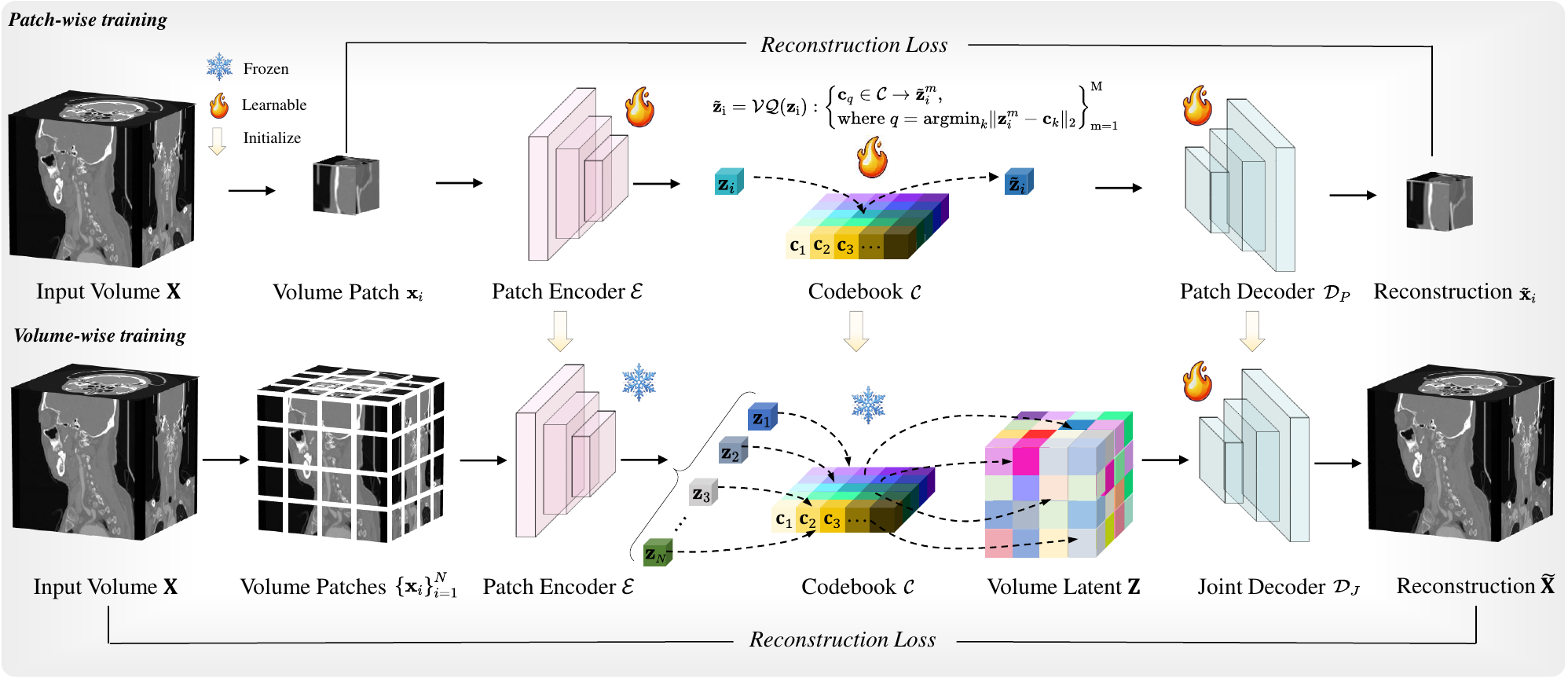}}
\caption{
Patch-Volume Autoencoder with a two-stage training strategy. 
In the first stage, the model is trained solely to compress and reconstruct small patches from high-resolution volumes. 
In the second stage, all parameters are fixed except for the decoder, which is fine-tuned on high-resolution volumes to become a joint decoder.
}
\label{fig:PVAE}
\vspace{-2mm}
\end{figure*}

\begin{figure*}[!t]
\centerline{\includegraphics[width=0.98\textwidth]{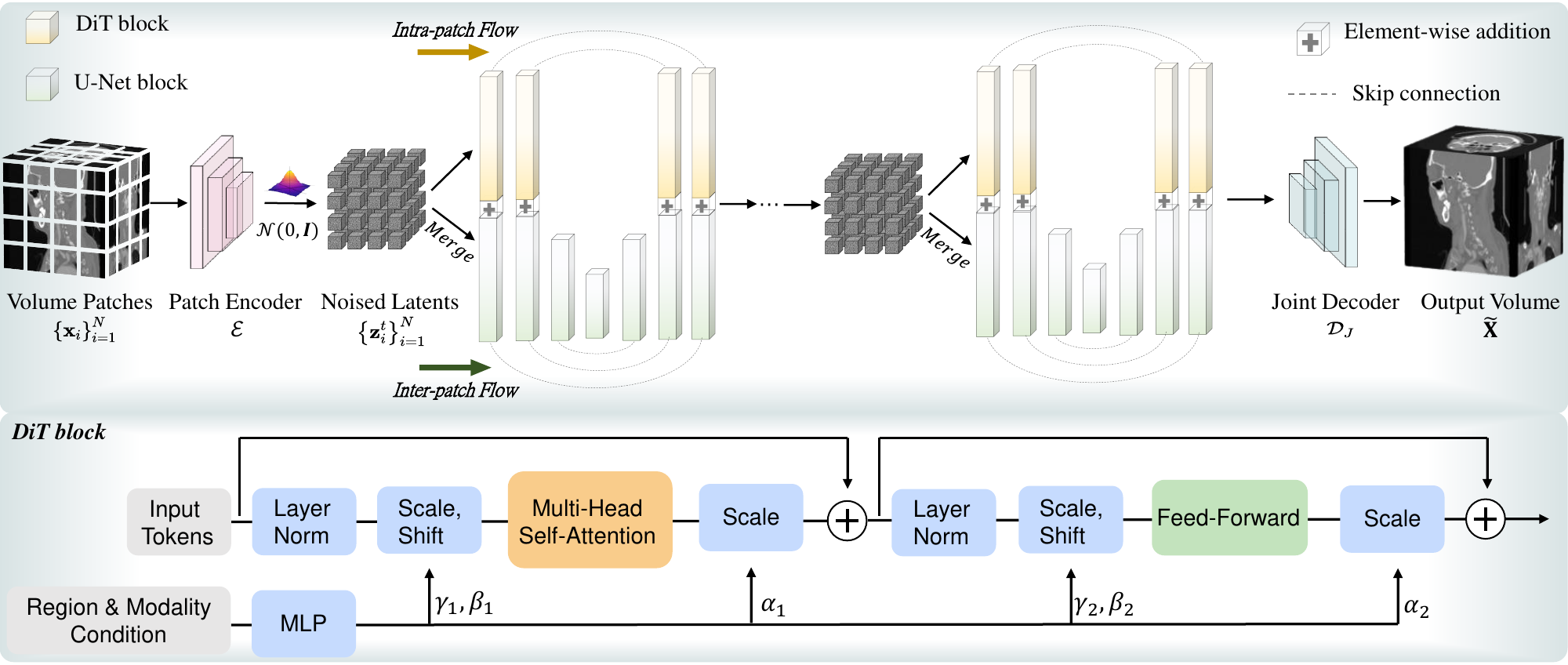}}
\caption{
BiFlowNet noise estimator. 
The intra-patch flow focuses on denoising each patch and recovering fine-grained local details, while the inter-patch flow is designed to capture and reconstruct the global structures across the entire volume.}
\label{fig:BiFlowNet}
\vspace{-4mm}
\end{figure*}

\subsection{Compact Latent Space for 3D Medical Image}
\label{sec:vqgan}

To enable efficient 3D medical image generation, we develop a novel compression framework that maps high-resolution 3D volumes into a compact, semantically meaningful latent space. 
It facilitates more efficient diffusion process.
Building upon VQ-VAE~\cite{van2017neural}, we introduce Patch-Volume Autoencoder with a two-stage training strategy to address the challenges of 3D medical image compression (\cref{fig:PVAE}), offering three key advantages. 
\textbf{(1) Memory efficient}. 
The first stage compresses and reconstructs small, fine-detailed patches, overcoming memory limitations of processing entire volumes.
\textbf{(2) Artifact-free reconstruction}. 
 The second stage fine-tunes the joint decoder, enabling seamless full-volume reconstruction without producing patch boundary artifacts.
\textbf{(3) Meaningful and compact representation}. 
The Vector Quantization (VQ) mechanism maps continuous representations to discrete codes in a learned codebook, capturing anatomical patterns and handling representation variations across anatomical regions.

\subsubsection{Patch-wise Training}
In the first stage, the 3D volume $\mathbf{X} \in \mathbb{R}^{H \times W \times D}$ is divided into smaller patches $\left \{ \mathbf{x}_i \in \mathbb{R}^{h \times w \times d}\right \}_{i=1}^{N}$, where $N$ denotes the total number of patches. 
We process these patches in a patch-wise manner using a patch encoder $\mathcal{E}$, a patch decoder $\mathcal{D}_P$, and a discrete and learnable codebook $\mathcal{C} = \{\mathbf{c}_k\in \mathbb{R}^{C}\}_{k=1}^{K}$ with $K$ denoting the codebook size

First, for each patch $\mathbf{x}_i$, the patch encoder $\mathcal{E}$ extracts the corresponding patch feature $\mathbf{z}_i=\mathcal{E}(\mathbf{x}_i)\in\mathbb{R}^{M \times C}$, where $M$ denotes the number of codes.
Second, $\mathbf{z}_i$ would then be transformed into quantized patch feature $\tilde{\mathbf{z}}_i\in\mathbb{R}^{M \times C}$ via vector quantization $\mathcal{VQ}$:

\begin{equation}
   \tilde{\mathbf{z}}_i = \mathcal{VQ}(\mathbf{z}_i): 
   \left\{
   \begin{aligned}
   &\mathbf{c}_q \in \mathcal{C} \rightarrow \tilde{\mathbf{z}}_i^m, \\
   &\mathrm{where}~q = \mathrm{argmin}_k \Vert \mathbf{z}_i^m - \mathbf{c}_k \Vert_2
   \end{aligned}
   \right\}_{m=1}^M.
\end{equation}
Third, the patch decoder $\mathcal{D}_P$ produces the patch reconstruction $\tilde{\mathbf{x}}_i=\mathcal{D}_P(\tilde{\mathbf{z}}_i)\in\mathbb{R}^{h\times w\times d}$ from the quantized patch feature $\tilde{\mathbf{z}}_i$.

\subsubsection{Volume-wise Training} 

We fine-tune the patch decoder $\mathcal{D}_P$ into a joint decoder $\mathcal{D}_J$, avoiding the need for individual patch processing and concatenation.
It enables high-resolution reconstruction devoid of boundary artifacts.

As shown in Fig.~\ref{fig:PVAE}, in this stage, the entire volume is first partitioned into smaller patches and each patch is then encoded and quantized to quantized patch features $\left \{ \tilde{\mathbf{z}}_i \right \}_{i=1}^{N}$, similarly to prior patch-wise training. 
Afterward, these quantized features are concatenated into a latent volume $\tilde{\mathbf{Z}}\in\mathbb{R}^{N\cdot M \times C}$, which is then decoded by the joint decoder $\mathcal{D}_J$ to produce the final reconstruction $\tilde{\mathbf{X}}=\mathcal{D}_J(\tilde{\mathbf{Z}})\in\mathbb{R}^{H\times W\times D}$ at once.
% Unlike patch-wise training, $\mathcal{D}_J$ processes the entire volume at once. 
Notably, the parameters of the patch encoder $\mathcal{E}$ and the codebook $\mathcal{C}$ are initialized from patch-wise training stage and kept frozen during this stage, while the joint decoder $\mathcal{D}_J$ is learnable. 
This approach is memory-efficient, as it only requires storing the backward gradients for the joint decoder, $\mathcal{D}_J$. To demonstrate this, we report memory usage in~\cref{tab:memory_comparison}, comparing ``Naive'' volume-wise training, where gradients for all components are stored, with our ``Efficient'' volume-wise training, where only gradients for the joint decoder $\mathcal{D}_J$ are stored. We tested three input sizes: ($64 \times 64 \times 64$), ($128 \times 128 \times 128$), and ($192 \times 192 \times 192$), and present the \textbf{memory usage per image} and \textbf{maximum batch size} that can be processed on a single 80GB GPU. \textcolor{red}{Note that we did not report results for ``Naive'' training with a ($192 \times 192 \times 192$) input size, as its memory consumption exceeds the available 80GB. Our two-stage training (patch-wise encoding followed by volume-wise integration learning) specifically enables the model to learn how to coherently assemble patch latent representations into larger volumes, even if full volumes of that final size (e.g., $256^3$ or $512^3$) were not trained end-to-end during the volume-wise training stage.}
\vspace{-2mm}

\begin{table}[htbp]
\centering
\captionsetup{justification=centering, labelsep=newline} 
\renewcommand{\arraystretch}{1.4} % Increase row spacing
\caption{Comparison of memory usage and maximum batch size for different image sizes with respect to two training strategies}

\begin{tabularx}{\linewidth}{>{\centering\arraybackslash}c|>{\centering\arraybackslash}X|>{\centering\arraybackslash}X|>{\centering\arraybackslash}X|>{\centering\arraybackslash}X}
\hline
Image size & \multicolumn{2}{c|}{Naive} & \multicolumn{2}{c}{Efficient} \\ \hline
           & Memory (GB) & Batch size & Memory (GB) & Batch size \\ \hline
$64\times 64 \times 64$ & 3.56 & 28 & 2.88 & 41 \\ \hline
$128\times 128 \times 128$ & 24.10 & 3 & 16.21 & 5 \\ \hline
$192\times 192 \times 192$ & / & / & 52.48 & 1 \\ \hline
\end{tabularx}
\label{tab:memory_comparison}
\end{table}

\subsubsection{Training Loss}
We use the same loss functions for both training stages. 
In this section, we use $\mathbf{x}$ to replace either $\mathbf{x}_i$ or $\mathbf{X}$ and omit the dimensionality for simplicity.
The full loss function for the Autoencoder, denoted as $\mathcal{L}_{AE}$, is formulated as follows:
\begin{equation}
\mathcal{L}_{AE} = \mathcal{L}_{VQ} + \lambda_{Adv}\mathcal{L}_{Adv} + \lambda_{TP}\mathcal{L}_{TP},
\end{equation}
where $\mathcal{L}_{VQ}$, $\mathcal{L}_{Adv}$, and $\mathcal{L}_{TP}$ are vector quantization loss, adversarial loss and tri-plane loss, respectively. 
$\lambda_{Adv}$ and $\lambda_{TP}$ are loss weights for $\mathcal{L}_{Adv}$ and $\mathcal{L}_{TP}$, respectively.
Each loss function is detailed in the following.

$\mathcal{L}_{VQ}$ is the fundamental loss for Patch-Volume Autoencoder training~\cite{van2017neural}, defined as:
\begin{align}
\mathcal{L}_{VQ} = \Vert \mathbf{x} - \tilde{\mathbf{x}} \Vert_2 &+ \Vert \mathrm{sg}[\mathcal{E}(\mathbf{x})]-\tilde{\mathbf{z}} \Vert_2 \notag \\
 &+ \Vert \mathrm{sg}[\tilde{\mathbf{z}}]-\mathcal{E}(\mathbf{x}) \Vert_2,
\end{align}
where $\tilde{\mathbf{x}}$ is the reconstruction counterpart of $\mathbf{x}$, $\tilde{\mathbf{z}}$ is the quantized feature, and $\mathrm{sg}[\cdot]$ stands for stop-gradient operation.
This loss function minimizes the disparity between input data $\mathbf{x}$ and its reconstruction $\tilde{\mathbf{x}}$, while aligning encoded features $\mathcal{E}(\mathbf{x})$ and feature codes $\tilde{\mathbf{z}}$ searched from codebook $\mathcal{C}$.

$\mathcal{L}_{Adv}$ incorporates a discriminator $\mathfrak{D}$ to encourage more realistic reconstructions~\cite{esser2021taming}:
\begin{equation}
\mathcal{L}_{Adv} = \log \mathfrak{D}(\mathbf{x}) + \log (1-\mathfrak{D}(\tilde{\mathbf{x}})).
\end{equation}

$\mathcal{L}_{TP}$ is employed to maintain high perceptual quality during vector quantization with help of perceptual loss~\cite{johnson2016perceptual,larsen2016autoencoding,lamb2016discriminative}.
In order to leverage existing 2D pre-trained feature extractor, we randomly select three orthogonal 2D planes from input data $\{\mathbf{p}_n\}_{n=1}^{3}\subset\mathbf{x}$ and reconstruct the planes $\{\tilde{\mathbf{p}}_n\}_{n=1}^{3}\subset\tilde{\mathbf{x}}$ at the same position to compute perceptural loss:
\begin{equation}
\mathcal{L}_{TP} = \sum_{n=1}^{3} \Vert \phi(\mathbf{p}_n) - \phi(\tilde{\mathbf{p}}_n)\Vert_2,
\end{equation}
where $\phi$ is the pre-trained VGG-16 network~\cite{simonyan2014very, deng2009imagenet}.

\subsection{3D Medical Diffusion Model}
\label{sec:dm}
\subsubsection{Diffusion Model}
The diffusion model in 3D MedDiffusion operates in the compact latent space, either patch-wise ($\mathbf{x}_i$) or volume-wise ($\mathbf{X}$).
In this section, we use $\mathbf{x}$ to denote both and omit dimensionality for simplicity.
Diffusion model first defines a forward noising process that incrementally adds noise to the initial latent $\mathbf{z}^{0}=\mathcal{E}(\mathbf{x})$, formulated as:

\begin{equation}
\begin{aligned}
q(\mathbf{z}^{1:T} \vert \mathbf{z}^{0}) &\coloneqq  \prod_{t=1}^{T} q(\mathbf{z}^t \vert \mathbf{z}^{t-1}), \\
\text{where} \quad q(\mathbf{z}^t \vert \mathbf{z}^{t-1}) &= \mathcal{N}(\mathbf{z}^t; \sqrt{1-\beta^t}\mathbf{z}^{t-1}, \beta^t\mathbf{I}).
\end{aligned}
\end{equation}
Here \( \sqrt{1 - \beta^t} \mathbf{z}^{t-1} \) is the mean and \( \beta^t \mathbf{I} \) is the covariance of the distribution at timestep \( t \in \{1, 2, \dots, T\} \), with \( T \) denoting the total number of timesteps. \( \mathbf{I} \) is the identity matrix, and \( \beta_t \in (0, 1) \) is a noise-level hyperparameter.

Using the reparameterization trick~\cite{rezende2015variational}, one can sample $\mathbf{z}^t$ as follows:
\begin{equation}
\mathbf{z}^t = \sqrt{\overline{\alpha}^t} \, \mathbf{z}^{0} + \sqrt{1 - \overline{\alpha}^t} \, \epsilon, \quad \epsilon \sim \mathcal{N}(0, \mathbf{I}),
\label{eq:7}
\end{equation}

where $\alpha^t = 1 - \beta^t$  and $\overline{\alpha}^t = \prod_{i=1}^{t} \alpha^i$. For the reverse process, the diffusion model is trained to learn a denoising network $\mu_{\theta}$, expressed as:
\begin{equation}
p_{\theta}(\mathbf{z}^{t-1} \vert \mathbf{z}^t) = \mathcal{N}(\mathbf{z}^{t-1}; \mu_{\theta}(\mathbf{z}^t, t), \sigma^t \mathbf{I}),
\label{eq:8}
\end{equation}

Here $\mu_{\theta}(\mathbf{z}^t, t)$ and $\sigma^t \mathbf{I}$ are the mean and covariance. By combining~\cref{eq:7} and~\cref{eq:8}, we can derive:
\begin{equation}
\mu_{\theta}(\mathbf{z}^t, t) = \frac{1}{\sqrt{\alpha_t}} \left( \mathbf{z}^t -\frac{\beta_t}{\sqrt{1 - \overline{\alpha}^t}}  \,  \, \epsilon_{\theta}(\mathbf{z}^t, t) \right).
\end{equation}

The training objective is to train a noise estimator $\epsilon_{\theta}$ to predict the added Gaussian noise $\boldsymbol{\epsilon}$, described by the loss function:
\begin{equation}
\mathcal{L}_{Diff} = \mathbb{E}_{\mathbf{z}^0, \boldsymbol{\epsilon}, t, \mathbf{c}}\Big [ \left \| \boldsymbol{\epsilon} -  \epsilon _{\theta}(\mathbf{z}^t,t,\mathbf{c}) \right \|_2  \Big ]  ,
\end{equation}
where $\mathbf{c}$ is the class condition, indicating the data modality and anatomical regions. 
Once $p_{\theta}(\mathbf{z}^{t-1} \vert \mathbf{z}^t)$ is trained, new latents can be generated by progressively sampling $\mathbf{z}^{t-1} \sim p_{\theta}(\mathbf{z}^{t-1} \vert \mathbf{z}^t)$.

\subsubsection{BiFlowNet Architecture}
We propose a novel noise estimator called BiFlowNet, which integrates both intra-patch flow and inter-patch flow to generate latent representations. 
Intra-patch flow captures fine-grained details, whereas inter-patch flow maintains global structural consistency.
The dual-flow latent features are integrated to achieve both objectives simultaneously.

% It consists of three key components, including intra-patch flow diffusion process, inter-patch flow diffusion process, and dual flow integration, as detailed in following.

\textbf{Intra-Patch Flow.}
In our approach, high-resolution 3D medical images are encoded into latent space on patch-wise basis, resulting in each patch following an independent distribution.
Intra-patch flow focuses on denoising each patch and recovering fine-grained local details. 
We adopt diffusion transformer (DiT)~\cite{peebles2023scalable} as the backbone for intra-patch flow, leveraging its exceptional performance in complex generative tasks on large datasets.
Class embedding is also incorporated as input to enable multi-class training for different regions and modalities.
The DiT block architecture follows the design in~\cite{peebles2023scalable}, with adaptations for 3D inputs such as expanded spatial dimensions, as shown in~\cref{fig:BiFlowNet}.

\textbf{Inter-Patch Flow.}
However, generating only patch-wise regions fails to preserve global structural consistency when reconstructing the entire 3D volume.
In response to this challenge, we design inter-patch flow to capture and reconstruct the global structures across the entire volume.
We opt a standard 3D U-Net~\cite{ronneberger2015u} as the backbone to eliminate computational overhead raised from high-resolution 3D images.

\textbf{Dual Flow Integration.}
As shown in~\cref{fig:BiFlowNet}, to avoid isolated optimization, we combine local and global features from dual flows with element-wise addition at each timestep.
Features from DiT's first and last two blocks are merged with U-Net's corresponding blocks.
The latent volume is then processed by the U-Net and divided into patches for the DiT at the next timestep.

\subsection{ControlNet for Downstream Tasks}
\label{sec:controlnet}
\begin{figure}[htbp]
\centerline{\includegraphics[width=0.98\linewidth]{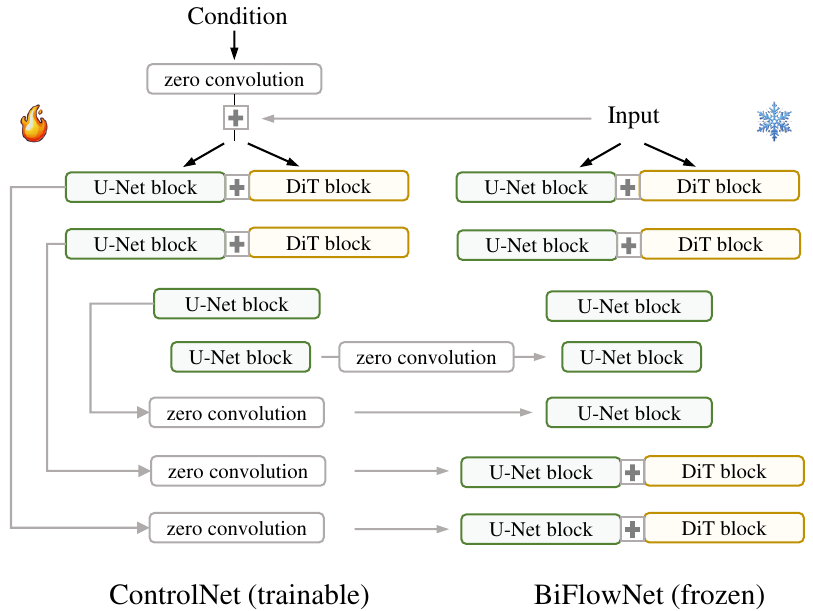}}
\caption{The network architecture of ControlNet}
\label{fig:controlnet}
\vspace{-2mm}
\end{figure}

To improve controllability and adaptability for diverse downstream tasks, we integrate ControlNet~\cite{zhang2023adding} into the diffusion model.
ControlNet is specifically designed to inject additional task-specific conditions into the diffusion model via fine-tuning. 
In this process, the parameters of the pre-trained diffusion model are frozen, while a cloned, trainable encoder of BiFlowNet is introduced. 
This trainable copy accepts task-specific condition as input and its output is connected to the pre-trained diffusion model via zero convolutions~\cite{zhang2023adding}. 
The network architecture is shown in~\cref{fig:controlnet}.
Since the diffusion model in 3D MedDiffusion operates within the latent space, it is essential to encode the task-specific condition into this space, denoted as $\mathbf{c}_{task} $.
The loss function for the fine-tuning process is defined as:
\begin{equation}
\mathcal{L}_{Con} = \mathbb{E}_{\mathbf{z}^0, \boldsymbol{\epsilon}, t, \mathbf{c}, \mathbf{c}_{task}}\Big [ \vert \vert \boldsymbol{\epsilon} -  \epsilon_{\theta}(\mathbf{z}^t,t,\mathbf{c},\mathbf{c}_{task}) \vert \vert _2 \Big ].
\label{eq:controlnet}
\end{equation}
This control mechanism allows the 3D MedDiffusion to evolve from a generic generative model into a task-specific model by leveraging the pre-trained model's strong prior knowledge.

Moreover, the fine-tuning process is notably time-efficient, as only a small set of parameters require training.

\begin{table}[htbp]
\centering
\captionsetup{justification=centering, labelsep=newline} 
\renewcommand{\arraystretch}{1.4} % Increase row spacing
\caption{Training and inference time of each comparison methods on the \textbf{CTChestAbdomen} dataset}
\begin{tabularx}{\linewidth}{>{\centering\arraybackslash}c|>{\centering\arraybackslash}X|>{\centering\arraybackslash}X}
\hline
                 & \textbf{Training Time} & \textbf{Inference Time} \\ \hline\hline
HA-GAN           & \textcolor{red}{30h}                   & \textless 1s            \\ \hline
MedicalDiffusion & 79h                   & 83s                     \\ \hline
WDM              & \textcolor{red}{210h}                   & 251s                    \\ \hline
MAISI            & 81h                   & 82s                     \\ \hline
MedSyn           & 71h                   & 255s                    \\ \hline
Ours (CT)        & 83h                   & 128s                    \\ \hline
\end{tabularx}
\label{tab:time}
\end{table}

\vspace{-5mm}

%%Experiments...........
\section{Experiments}
In this section, we begin by detailing the datasets (\cref{sec:Dataset}) and implementation details (\cref{sec:Implementation Details}). 
Next, we compare our 3D MedDiffusion with competing methods (\cref{sec:Comparison}). 
Finally, we assess the generalizability of 3D MedDiffusion across four downstream tasks: sparse-view CT reconstruction, MRI reconstruction, data augmentation for segmentation and data augmentation for classification (\cref{sec:Downstream}).

\subsection{Experimental Setting}

\subsubsection{Dataset}
\label{sec:Dataset}
To develop a large-scale 3D generative model, we train the 3D MedDiffusion on an extensive collection of high-quality 3D medical image datasets.
These datasets include CT scans of head and neck, chest and abdomen, and lower limb, as well as MR images of brain, chest, abdomen, and knee, covering the major anatomical regions of the human body.
We organize these multi-modality and region-mixed datasets into six sub-datasets: \textbf{CTHeadNeck}, \textbf{CTChestAbdomen}, \textbf{CTLowerLimb}, \textbf{MRBrain}, \textbf{MRChestAbdomen}, \textbf{MRKnee}. 

\textbf{CTHeadNeck Dataset.}
For head and neck region of CT modality, we collect one public dataset and two in-house datasets.
The public dataset is the CQ500 dataset~\cite{chilamkurthy2018development} with 491 CT scans.
The in-house datasets consist of a CTA dataset with 256 images and a CBCT dataset with 600 images.
In total, we collect 1,347 images with image size ranging from $256 \times 256 \times 128$ to $256 \times 256 \times 256$ and voxel spacing below 1.25$\mathrm{mm}$ along all three axes.

\textbf{CTChestAbdomen Dataset.}
For chest and abdomen region of CT modality, we involve four public CT datasets: 
The Pulmonary Airway dataset~\cite{wang2021medical,zheng2021alleviating,yu2022break,qin2019airwaynet} with 300 CT scans.
% on the pulmonary region.
The LIDC-IDRI dataset~\cite{armato2011lung} with 1,010 CT scans.
The CTSpline1K dataset~\cite{deng2021ctspine1k} with 1,005 CT scans.
The AbdomenCT-1K dataset~\cite{ma2021abdomenct} with 1,112 CT scans.
In all these four datasets, we gather 3,000 images after excluding low-quality ones with image size ranging from $256 \times 256 \times 256$ to $512 \times 512 \times 512$ and voxel spacing below 1.25$\mathrm{mm}$ along all three axes.

\textbf{CTLowerLimb Dataset.}
For lower limb region of CT modality, we develop an in-house dataset with 400 CT scans, covering the area from the thigh to the foot.
The image size ranges from $256 \times 256 \times 256$ to $256 \times 256 \times 512$, with voxel spacing below 1.25$\mathrm{mm}$ along three axes.

\textbf{MRBrain Dataset.}
For brain region of MR modality, we collect 3,000 MR images from the UK Biobank~\cite{sudlow2015uk}, comprising 1,500 T1-weighted images and 1,500 T2-weighted FLAIR images.
Non-brain regions have been excluded to enhance model focus.
All images have a size of $192 \times 192 \times 192$ and a voxel spacing of 1$\mathrm{mm}$ along all three axes.

\textbf{MRChestAbdomen Dataset.}
For chest and abdomen region of MR modality, we create an in-house dataset comprising 1,826 MR images, including both T1-weighted and T2-weighted data.
All images have a size of $256 \times 256 \times 128$ and a voxel spacing of 1.25$\mathrm{mm}$ along the $x$- and $y$-axes, and 2$\mathrm{mm}$ along the $z$-axis.

\textbf{MRKnee Dataset.}
For knee region of MR modality, we use the publicly available fastMRI~\cite{zbontar2018fastmri} dataset, focusing exclusively on single-coil data for model development. 
This dataset comprises 1172 volumes, each consisting of approximately 40 slices with size of $320 \times 320$.
We interpolate the data along the $z$-axis, standardizing each volume to a size of $320 \times 320 \times 64$. 
All images have a voxel spacing of 1$\mathrm{mm}$ along the $x$- and $y$-axes, and 5$\mathrm{mm}$ along the $z$-axis.

During the data collection process, for some public datasets, we perform random selection to choose an appropriate amount of data from the original dataset to ensure class balance. For example, we use only 3,000 MR brain images from the original dataset. Additionally, we exclude images of low quality, such as those with very large spacing or abnormal FOV (Field of View). We perform the max-min normalization, scaling the voxel values to the range of [-1, 1].

\subsubsection{Competing Methods}
\label{sec:Competing Methods}
We select several generative methods as baselines, including both GAN-based and diffusion-based approaches, as detailed below.
HA-GAN~\cite{sun2022hierarchical} is a hierarchical amortized GAN which generates low-resolution images and high-resolution sub-volumes simultaneously to address memory constraints.
Medical Diffusion~\cite{khader2022medical} is a video diffusion model trained in the learned latent space of a 3D VQ-GAN.
WDM~\cite{friedrich2024wdm} is a diffusion-based framework for medical image synthesis using wavelet-decomposed images.
MAISI~\cite{guo2024maisi} is a 3D U-Net trained in the learned latent space of a 3D VAE. 
MedSyn~\cite{xu2024medsyn} is a diffusion-based framework designed for generating high-fidelity 3D CT images with textual guidance.

\subsubsection{Comparison Settings}
To ensure a fair comparison, we trained three models for our method: Ours (CT), trained exclusively on \textbf{CTChestAbdomen} dataset, Ours (MR), trained exclusively on \textbf{MRBrain}, and Ours (universal) trained on all of the six dataset mentioned above. Then we retrained all of the competing methods listed in~\cref{sec:Competing Methods} on \textbf{CTChestAbdomen} dataset for comparison with Ours (CT), and also on \textbf{MRBrain} for comparison with Ours (MR).

\subsubsection{Implementation Details}  
\label{sec:Implementation Details}
For Patch-Volume Autoencoder, during the patch-wise training phase, we set the patch size to $64 \times 64 \times 64$, with a compression ratio of $4 \times 4 \times 4$. 
The codebook contains 8192 codes with a dimensionality of 8. 
The learning rate is $3 \times 10^{-4}$. 
We start to train the discriminator after 20,000 iterations. 
In the volume-wise training phase, only the decoder and discriminator are trainable.
The learning rate is lowered to $3 \times 10^{-5}$.
We set $\lambda_{Adv} = 2$ and $\lambda_{TP} = 4$ for both phases.

For BiFlowNet noise estimator, we employ the cosine noise schedule~\cite{nichol2021improved} with $T=1000$ timesteps, and the learning rate of $1 \times 10^{-4}$ with polynomial learning rate decay.

In the comparison experiments, we trained all of the comparison methods from scratch including Ours (CT) and Ours (MR) under the same hardware condition, i.e., a single NVIDIA A100 80 GB GPU. The training and inference times for each competing method on the \textbf{CTChestAbdomen} dataset are reported in~\cref{tab:time}. \textcolor{red}{For training time, all methods were trained for over 10 days. We selected the best checkpoint to report both training time and evaluation metrics.} For the inference time, we performed inference for all methods on generating a single image at $256 \times 256 \times 256$ resolution.

Moreover, for Ours (universal), the training of the Patch-Volume Autoencoder for this model was performed on a single NVIDIA A100 80 GB GPU, requiring 72 GPU hours. The training for the BiFlowNet noise estimator, 8 NVIDIA A100 80 GB GPUs were used, requiring a total of 120 GPU hours.

Regarding resolution, all competing methods, Ours (CT), and Ours (MR) were trained using a uniform resolution by resizing all images in the \textbf{CTChestAbdomen} dataset to $256 \times 256 \times 256$, while images in the \textbf{MRBrain} dataset were not resized, as they are originally at $192 \times 192 \times 192$ size. However, for Ours (universal), uniform resolution was not applied.

\subsubsection{Evaluation Metrics}
We evaluate the fidelity and diversity of generated images using Fréchet Inception Distance (FID)~\cite{heusel2017gans} and Maximum Mean Discrepancy (MMD)~\cite{gretton2012kernel}, which assess the similarity between the distributions of real and generated images.
For feature extraction in FID and MMD calculations, we use a 3D ResNet model pre-trained on 3D medical images~\cite{chen2019med3d}.
Lower FID and MMD values indicate higher realism of the generated images.
Besides, we use the Multi-Scale Structural Similarity Index (MS-SSIM) to assess the diversity of the generated images. 
MS-SSIM measures the structural similarity between generated images across multiple scales, with lower values reflecting higher diversity.

\subsection{Comparison}
\label{sec:Comparison}

\subsubsection{Quantitative Evaluation}

We generated 1,000 images for each competing method for quantitative evaluation. The quantitative experimental results for \textbf{CTChestAbdomen} dataset are provided in~\cref{tab:CTGen}.
For fairness, we exclude the results of Ours (universal) from the following analysis, as it is trained on more extensive resources and possesses a stronger capability, although its results are still presented.
Ours (CT) achieves the lowest FID and MMD scores compared with other competing methods, demonstrating superior generative fidelity and realism.
% In contrast, HA-GAN, MedicalDiffusion, and WDM struggles to generate high-resolution medical images. 
Compared to the second-best method, MAISI,  FID and MMD scores for Ours (CT) are over two times lower, highlighting a significant improvement in generative fidelity.
While MedicalDiffusion achieves the best performance in MS-SSIM metric, our method obtains comparable results.
The quantitative experimental results for the \textbf{MRbrain} dataset are summarized in~\cref{tab:MRGen}. 
Our method also achieves the best performance among all the metrics by a considerable margin.

\begin{table}[t]
\centering
\captionsetup{justification=centering, labelsep=newline} 
\renewcommand{\arraystretch}{1.7} % 调整行间距
\caption{Quantitative results on \textbf{CTChestAbdomen}}
\begin{tabularx}{\linewidth}{>{\centering\arraybackslash}c|>{\centering\arraybackslash}X|>{\centering\arraybackslash}X|>{\centering\arraybackslash}X} % 添加竖线
\hline
                 & FID~$\downarrow$      & MS-SSIM~$\downarrow$            & MMD~$\downarrow$               \\ \hline\hline
HA-GAN           & \textcolor{red}{0.0367}                     & \textcolor{red}{0.5723}                     & \textcolor{red}{1.0387}                     \\ \hline
MedicalDiffusion & 0.0284                     & 0.2227               & 0.6500                     \\ \hline
WDM              & \textcolor{red}{0.0203}                     & \textcolor{red}{0.2495}                     & \textcolor{red}{0.4656}                     \\ \hline
MAISI            & 0.0135                     & 0.2713                     & 0.2782                     \\ \hline
MedSyn           & 0.0174                     & 0.2725                &0.3122 \\ \hline
Ours(CT)          & 0.0055            & 0.2238                   &0.1049             \\ \hline
Ours (universal)            & \textbf{0.0041}            & \textbf{0.2173}     &\textbf{0.1023}             \\ \hline
\end{tabularx}
\label{tab:CTGen}
\end{table}

% \begin{table}[t]
% \centering
% \captionsetup{justification=centering, labelsep=newline} 
% \renewcommand{\arraystretch}{1.7} % 调整行间距
% \caption{Quantitative results on \textbf{CTChestAbdomen}}
% \begin{tabularx}{\linewidth}{>{\centering\arraybackslash}c|>{\centering\arraybackslash}X|>{\centering\arraybackslash}X|>{\centering\arraybackslash}X} % 添加竖线
% \hline
%                  & FID~$\downarrow$      & MS-SSIM~$\downarrow$            & MMD~$\downarrow$               \\ \hline\hline
% HA-GAN           & \textcolor{red}{0.0367}                     & \textcolor{red}{0.5723}                     & \textcolor{red}{1.0387}                     \\ \hline
% MedicalDiffusion & 0.0284                     & \textbf{0.2227}               & 0.6500                     \\ \hline
% WDM              & \textcolor{red}{0.0203}                     & \textcolor{red}{0.2495}                     & \textcolor{red}{0.4656}                     \\ \hline
% MAISI            & 0.0135                     & 0.2713                     & 0.2782                     \\ \hline
% MedSyn           & 0.0174                     & 0.2725                &0.3122 \\ \hline
% Ours(CT)          & \textbf{0.0055}            & 0.2238                   &\textbf{0.1049}             \\ \hline
% Ours (universal)            & \textbf{0.0041}            & \textbf{0.2173}     &\textbf{0.1023}             \\ \hline
% \end{tabularx}
% \label{tab:CTGen}
% \end{table}

\begin{table}[htbp]
\centering
\captionsetup{justification=centering, labelsep=newline} 
\renewcommand{\arraystretch}{1.7} % 调整行间距
\caption{Quantitative results on \textbf{MRBrain}}
\begin{tabularx}{\linewidth}{>{\centering\arraybackslash}c|>{\centering\arraybackslash}X|>{\centering\arraybackslash}X|>{\centering\arraybackslash}X} % 添加竖线
\hline
                 & FID~$\downarrow$      & MS-SSIM~$\downarrow$            & MMD~$\downarrow$               \\ \hline\hline
HA-GAN           & 0.1284                     & 0.7443                     & 4.4638                     \\ \hline
MedicalDiffusion & 0.0291                     & 0.7847                     & 1.9492                     \\ \hline
WDM              & 0.0057                     & 0.7844                     & 0.8742                     \\ \hline
MAISI            & 0.0075                     & 0.7380                     & 0.6957                     \\ \hline
MedSyn           & 0.0087    & 0.7778 & 0.7760                 \\ \hline
Ours(MR)             & 0.0044            & 0.7036                     & 0.6372                 \\ \hline
Ours (universal)             & \textbf{0.0039}            & \textbf{0.7027}                     & \textbf{0.6199}                 \\ \hline
\end{tabularx}
\label{tab:MRGen}
\end{table}

% \begin{table}[htbp]
% \centering
% \captionsetup{justification=centering, labelsep=newline} 
% \renewcommand{\arraystretch}{1.7} % 调整行间距
% \caption{Quantitative results on \textbf{MRBrain}}
% \begin{tabularx}{\linewidth}{>{\centering\arraybackslash}c|>{\centering\arraybackslash}X|>{\centering\arraybackslash}X|>{\centering\arraybackslash}X} % 添加竖线
% \hline
%                  & FID~$\downarrow$      & MS-SSIM~$\downarrow$            & MMD~$\downarrow$               \\ \hline\hline
% HA-GAN           & 0.1284                     & 0.7443                     & 4.4638                     \\ \hline
% MedicalDiffusion & 0.0291                     & 0.7847                     & 1.9492                     \\ \hline
% WDM              & 0.0057                     & 0.7844                     & 0.8742                     \\ \hline
% MAISI            & 0.0075                     & 0.7380                     & 0.6957                     \\ \hline
% MedSyn           & 0.0087    & 0.7778 & 0.7760                 \\ \hline
% Ours(MR)             & \textbf{0.0044}            & \textbf{0.7036}                     & \textbf{0.6372}                 \\ \hline
% Ours (universal)             & \textbf{0.0039}            & \textbf{0.7027}                     & \textbf{0.6199}                 \\ \hline
% \end{tabularx}
% \label{tab:MRGen}
% \end{table}

\begin{figure*}[t]
\centerline{\includegraphics[width=0.98\textwidth]{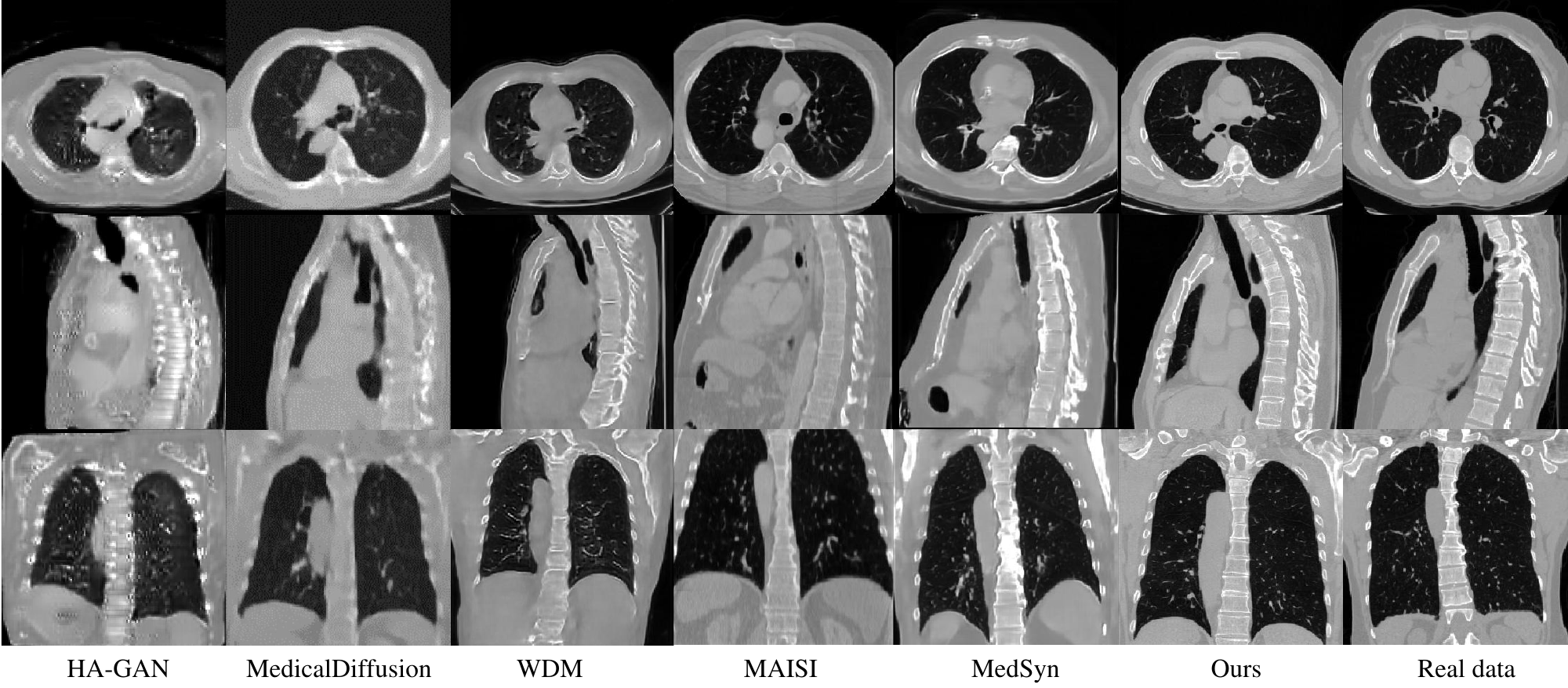}}
\caption{Qualitative results on \textbf{CTChestAbdomen}.  Window: [-1000,1000] HU.}

\label{fig:CTGen}
\vspace{-2mm}
\end{figure*}

\begin{figure*}[t]
\centerline{\includegraphics[width=0.98\textwidth]{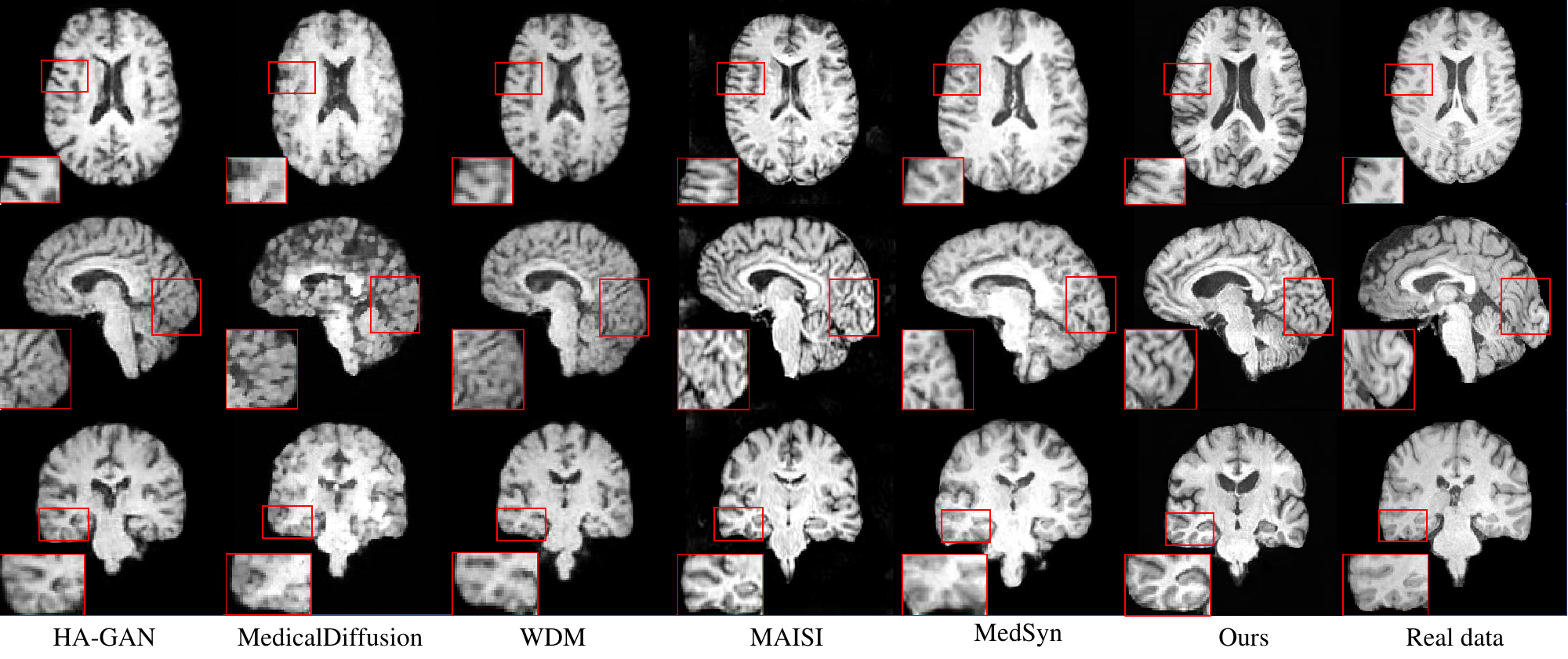}}
\caption{
Qualitative results on \textbf{MRBrain}.}
\label{fig:MRIGen}
\vspace{-4mm}
\end{figure*}

\subsubsection{Qualitative Evaluation}

To qualitatively evaluate the generation capabilities of all methods, we present some generated samples from \textbf{CTChestAbdomen} dataset (\cref{fig:CTGen}) and \textbf{MRBrain} dataset (\cref{fig:MRIGen}).
It is evident that HA-GAN, MedicalDiffusion, and WDM struggle to produce high-resolution CT images, resulting in noisy and blurry outputs with insufficient details.
MAISI and MedSyn produce satisfactory results but are slightly inferior to our method.
In contrast, our proposed method excels in capturing fine details like vertebrae in CT images.
It also produces sharp edges, such as the brain surface of MR images.

\begin{figure}[t]
\centerline{\includegraphics[width=0.98\linewidth]{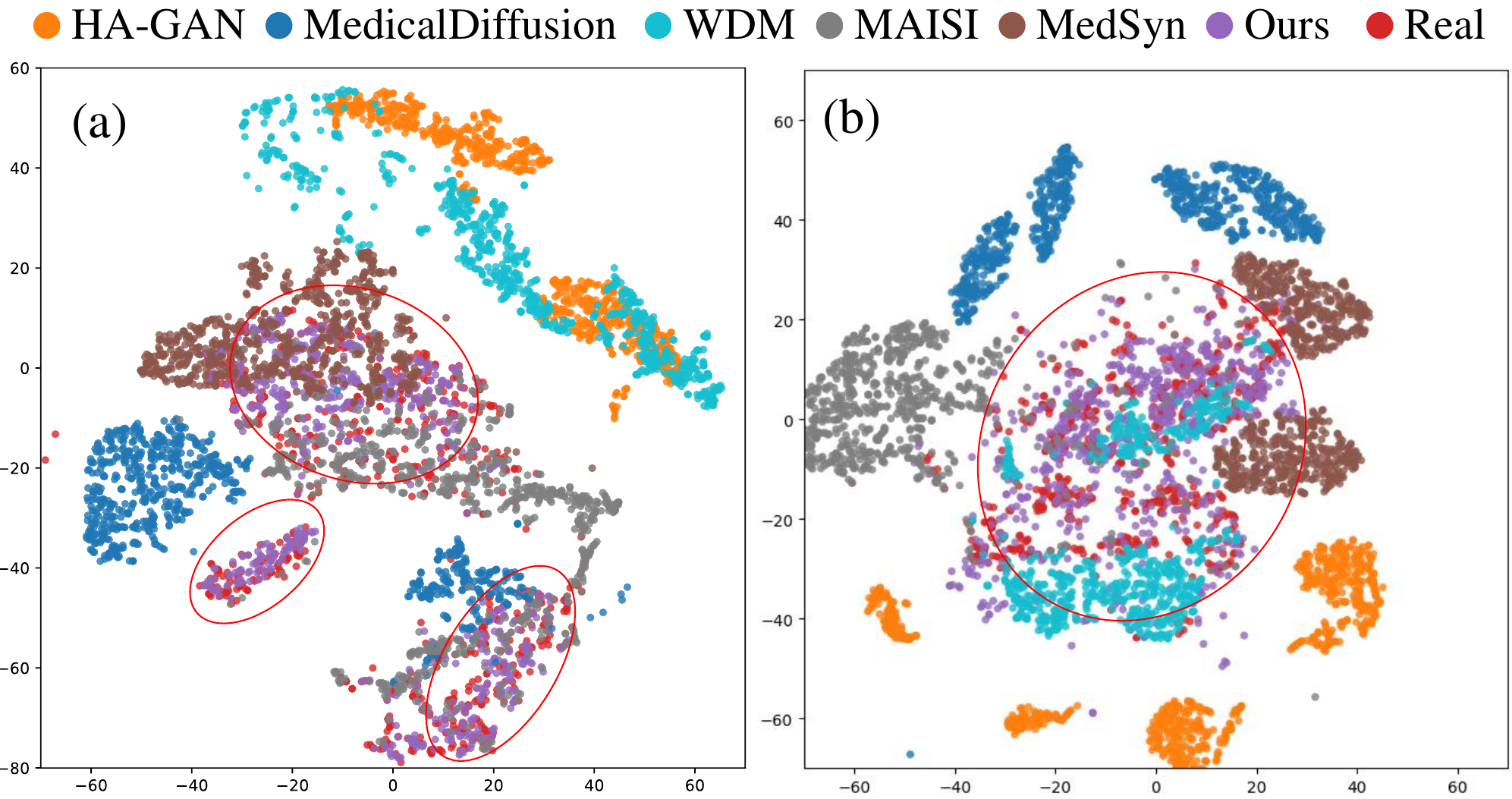}}
\caption{
t-SNE visualization on (a) \textbf{CTChestAbdomen} dataset and (b) \textbf{MRBrain} dataset.
Features extracted from the generated images are embedded into a 2-dimensional space. }

\label{fig:tSNE}

\end{figure}

To further evaluate the authenticity of the generated images, we project the images into a high-dimensional embedding space and apply a dimensionality reduction technique to visualize the distributions.
For comparison, we first employ a 3D pre-trained ResNet model~\cite{chen2019med3d} to extract features from 750 generated CT images for each method.
We then apply t-SNE~\cite{van2008visualizing} to reduce the dimensionality to 2.
This enables us to evaluate whether the generated distributions align well with real distributions in latent space. 
The resulting two-dimensional space is visualized in~\cref{fig:tSNE}. 
We observe that our generated distribution (purple dots) aligns more closely and compactly with the real distribution (red dots), as highlighted by the red ellipses. 
This demonstrates the superiority of our method in generating realistic images.

\subsection{Ablation Study}

In this section, we validate the effectiveness of two main components employed in our method, including Patch-Volume Autoencoder and BiFlowNet noise estimator.

\subsubsection{Patch-Volume Autoencoder}
Our method uses a two-stage training strategy for Autoencoder. In the first stage, entire network is trained in a patch-wise manner. In the second stage, the patch decoder is fine-tuned into a joint decoder in a volume-wise manner. To evaluate the effectiveness of our joint decoder and the uesed losses, we conducted two ablation studies.

In \textbf{Study a.}, we assess the effectiveness of our joint decoder. Since training the model in a purely volume-wise manner is not feasible due to memory constraints, we focus on evaluating the joint decoder by comparing it to the patch-wise decoder for image compression and reconstruction.

In \textbf{Study b.}, we investigate how the adversarial loss and tri-plane loss affect self-reconstruction performance. We compare four different settings: (1) removing the adversarial loss, $\mathcal{L}_{Adv}$, (2) removing the tri-plane loss, $\mathcal{L}_{TP}$, (3) removing both the adversarial and tri-plane losses, and (4) using the full loss.

Both \textbf{Study a.} and \textbf{Study b.} were conducted using the \textbf{CTChestAbdomen} dataset, with all settings controlled under the same hardware conditions. We evaluated performance using Peak Signal-to-Noise Ratio (PSNR) and Structural Similarity Index (SSIM)~\cite{SSIM} between the original and reconstructed images.

\begin{figure}[t]
\centerline{\includegraphics[width=0.98\linewidth]{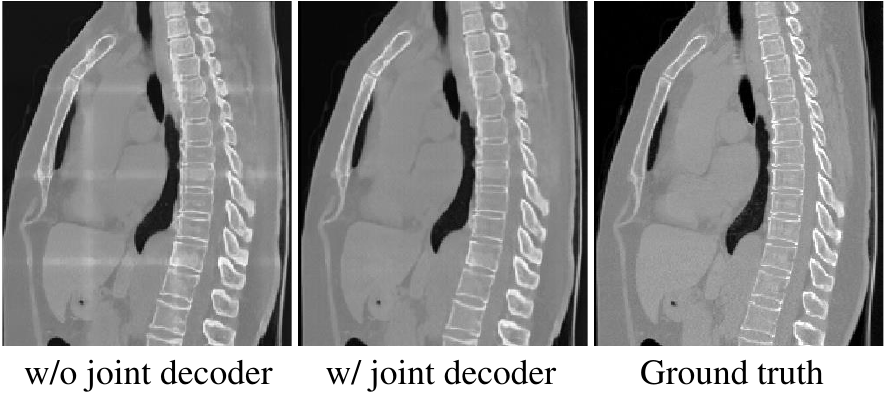}}
\caption{
Self-reconstruction results of ablation study on joint decoder. 
Window: [-1000, 1000] HU. 
}
\label{fig:ab_jointencoder}
\vspace{-2mm}
\end{figure}

The results of \textbf{Study a.} are provided in~\cref{tab:ab_joint_encoder}. Numbers outside parentheses represent the mean values, while those inside denote the standard deviations. The results indicate that incorporating the joint decoder significantly improves reconstruction quality across both metrics. Specifically, as shown in~\cref{fig:ab_jointencoder}, the joint decoder effectively eliminates patch boundary artifacts, which are commonly introduced by patch-wise decoding.

The results of \textbf{Study b.} are presented in~\cref{tab:ab_ae_loss}. Numbers outside parentheses represent the mean values, while those inside denote the standard deviations. When either $\mathcal{L}_{Adv}$ or $\mathcal{L}_{TP}$ is removed, performance drops significantly. Notably, the tri-plane loss, $\mathcal{L}_{TP}$, plays a critical role in improving performance, highlighting that the perceptual loss is highly effective for the self-reconstruction task.

\begin{figure}[t!]
\centerline{\includegraphics[width=0.98\linewidth]{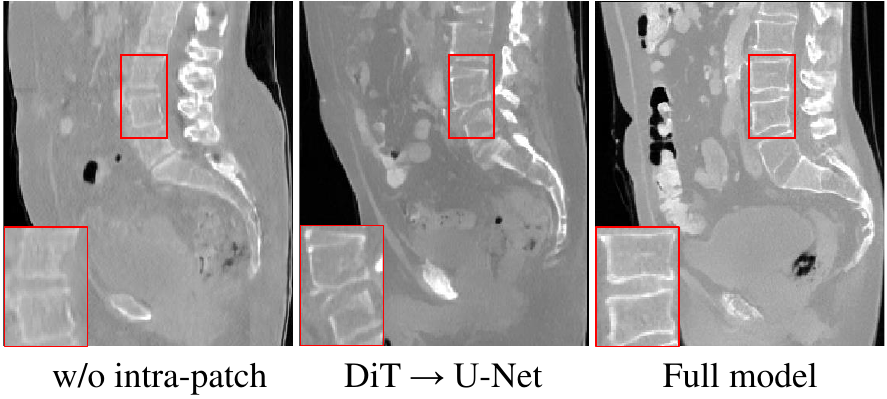}}
\caption{
Qualitative results of ablation study on BiFlowNet. Window:[-1000,1000] HU.
}
\label{fig:ab_biflownet}
\vspace{-4mm}
\end{figure}

\begin{table}[t]
\centering
\captionsetup{justification=centering, labelsep=newline} 
\renewcommand{\arraystretch}{1.4} % Increase row spacing
\caption{Quantitative results of ablation study on joint decoder}
\begin{tabularx}{\linewidth}{>{\centering\arraybackslash}c|>{\centering\arraybackslash}X|>{\centering\arraybackslash}X} % Center-align columns
\hline
                 & PSNR ($\mathrm{dB}$)~$\uparrow$ & SSIM~$\uparrow$\\ \hline\hline
w/o joint decoder   & 34.2307 (1.2492)  & 0.9273 (0.0925)  \\ \hline
w/ joint decoder    & 35.7916 (1.2187)  & 0.9314 (0.0963) \\ \hline
\end{tabularx}
\label{tab:ab_joint_encoder}
% \vspace{-3mm}
\end{table}

\begin{table}[t]
\centering
\captionsetup{justification=centering, labelsep=newline} 
\renewcommand{\arraystretch}{1.4} % Increase row spacing
\caption{Quantitative results of the ablation study on the loss function of Patch-Volume Autoencoder}
\begin{tabularx}{\linewidth}{>{\centering\arraybackslash}c>{\centering\arraybackslash}c>{\centering\arraybackslash}c|>{\centering\arraybackslash}X|>{\centering\arraybackslash}X} % Center-align columns
\hline
$\mathcal{L}_{VQ}$ & $\mathcal{L}_{Adv}$ & $\mathcal{L}_{TP}$ & PSNR ($\mathrm{dB}$)~$\uparrow$ & SSIM~$\uparrow$\\ \hline\hline
$\checkmark$       &                      &                      & 32.8993 (1.4932)   & 0.9172 (0.1092) \\ \hline              
$\checkmark$       & $\checkmark$         &                      & 33.2713 (1.2983)   & 0.9166 (0.0991) \\ \hline      
$\checkmark$       &                      & $\checkmark$        & 34.2783 (1.0423)   & 0.9184 (0.0954) \\ \hline   
$\checkmark$       & $\checkmark$         & $\checkmark$        & 35.7916 (1.2187)   & 0.9314 (0.0963) \\ \hline   
\end{tabularx}
\label{tab:ab_ae_loss}
\end{table}

\subsubsection{BiFlowNet}
To verify the effectiveness of our BiFlowNet noise estimator, we conducted two ablation studies: \textbf{Study a.}, focusing on the effectiveness of the architecture design. And \textbf{Study b.}, focusing on the compatibility with our proposed Autoencoder.

In \textbf{Study a.}, we compare two different settings: (1) removing the intra-patch flow from the noise estimator, and (2) replacing the DiT block of the intra-patch flow with a standard U-Net block. Note that the model cannot be trained with intra-patch flow alone, as the inter-patch flow is essential for the model to produce meaningful reconstructions across the entire volume.

In \textbf{Study b.}, to demonstrate that BiFlowNet is more suitable for patch-wise encoded latents and fully compatible with our proposed Autoencoder, we compare two settings: (1) training BiFlowNet on volume-wise encoded latents and (2) training BiFlowNet on patch-wise encoded latents. Our Patch-Volume Autoencoder employs patch-wise encoding and volume-wise decoding during training. Due to memory constraints, volume-wise encoding is not feasible in the training phase. However, during inference, we can perform volume-wise encoding to generate latents, as storing gradients is no longer necessary.

Both \textbf{Study a.} and \textbf{Study b.} were conducted using the \textbf{CTChestAbdomen} dataset, and all settings were controlled under the same hardware conditions. We evaluated performance using FID and MMD metrics.

The results of \textbf{Study a.} are presented in~\cref{tab:ab_biflownet} and~\cref{fig:ab_biflownet}. These findings clearly demonstrate that capturing intra-patch information is crucial for generating fine-grained local details. The DiT block outperforms the U-Net block, especially for generative models trained on large-scale datasets.

The results of \textbf{Study b.} are presented in~\cref{tab:ab_biflownet_R}. 
The findings indicate that, when training the noise estimator on patch-wise encoded latents, its performance is significantly better than the case when trained on volume-wise encoded latents. 
This highlights the perfect compatibility of our BiFlowNet noise estimator with our proposed Patch-Volume Autoencoder.

\begin{table}[t]
\centering
\captionsetup{justification=centering, labelsep=newline} 
\renewcommand{\arraystretch}{1.4} % Increase row spacing
\caption{Quantitative results of the ablation study on the BiFlowNet architecture}
\begin{tabularx}{\linewidth}{>{\centering\arraybackslash}c|>{\centering\arraybackslash}X|>{\centering\arraybackslash}X|>{\centering\arraybackslash}X} % Center-align columns
\hline
                 &w/o intra-patch  &DiT $\rightarrow$ U-Net &Full model\\ \hline\hline
FID~$\downarrow$   & 0.0117  & 0.0087  & 0.0055 \\ \hline
MMD~$\downarrow$    & 0.2392 & 0.1253 & 0.1049 \\ \hline
\end{tabularx}
\label{tab:ab_biflownet}
\vspace{-2mm}
\end{table}

\begin{table}[htbp]
\centering
\captionsetup{justification=centering, labelsep=newline} 
\renewcommand{\arraystretch}{1.4} % Increase row spacing
\caption{Quantitative results of ablation study on the encoding strategies}
\begin{tabularx}{\linewidth}{>{\centering\arraybackslash}c|>{\centering\arraybackslash}X|>{\centering\arraybackslash}X} % Center-align columns
\hline
                 & Volume-wise encoding  & Patch-wise encoding \\ \hline\hline
FID~$\downarrow$    & 0.0073  & 0.0055 \\ \hline
MMD~$\downarrow$    & 0.1147 & 0.1049 \\ \hline
\end{tabularx}
\label{tab:ab_biflownet_R}
\vspace{-5mm}
\end{table}

\subsection{Human Study}
\label{sec:HumanStudy}
\begin{figure}[htbp]
\centering
\includegraphics[width=1.0\columnwidth]{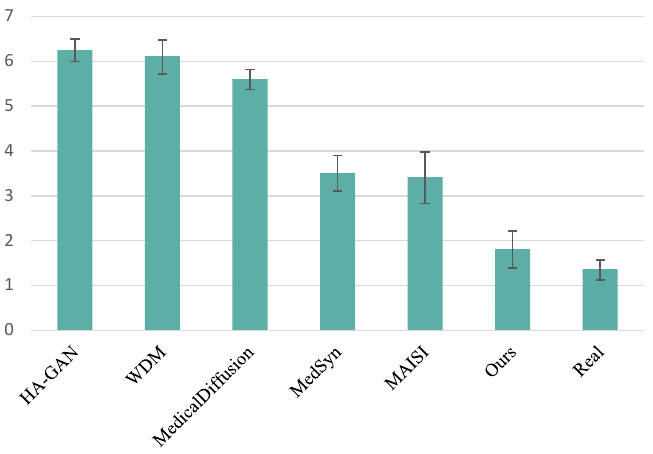}
\caption{The ranks for each method in human study.}
\label{figure:human_study}
\vspace{-4mm}
\end{figure}

To further evaluate the realism and quality of the generated images, we conducted a human study involving radiologists. 
This study was designed to assess both the perceived image quality and realism of the generated images from our proposed method, competing methods, and real images. 
This evaluation specifically focused on the chest and abdomen regions, which are critical areas in medical imaging.
A total of four radiologists participated in the evaluation process. 
Each radiologist was required to rank the images based on their overall visual quality and the degree to which they resembled real clinical images.
The images used in this human study were carefully selected to ensure that they provided a representative sample of typical clinical scenarios. 
Following the experimental setup described in~\cite{xu2024medsyn}, we designed an online survey that allowed the radiologists to rank the images in terms of quality and realism. 
The ranking scale used in the survey ranged from 1 to 7, with 1 representing the highest realism/quality and 7 representing the lowest. 
The mean ranks for each method are summarized in~\cref{figure:human_study}, which reveal that the mean rank of the images generated by our method closely aligns with the real images, outperforming the other methods in both quality and realism.

\begin{figure*}[ht!]
\centerline{\includegraphics[width=0.98\linewidth]{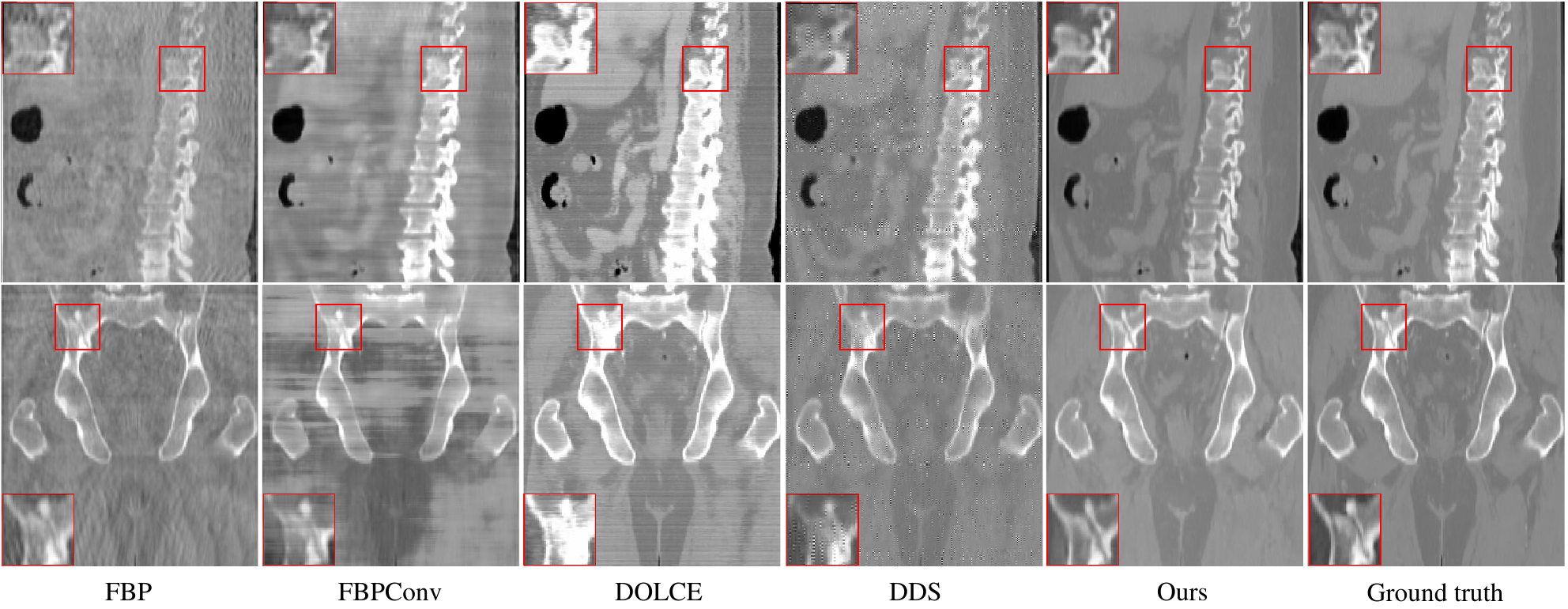}}
\caption{Qualitative results of sparse-view CT reconstruction. Window:[-1000,1000] HU.}
\label{fig:SVCT}
\vspace{-2mm}
\end{figure*}

\begin{figure*}[ht!]
\centerline{\includegraphics[width=0.98\textwidth]{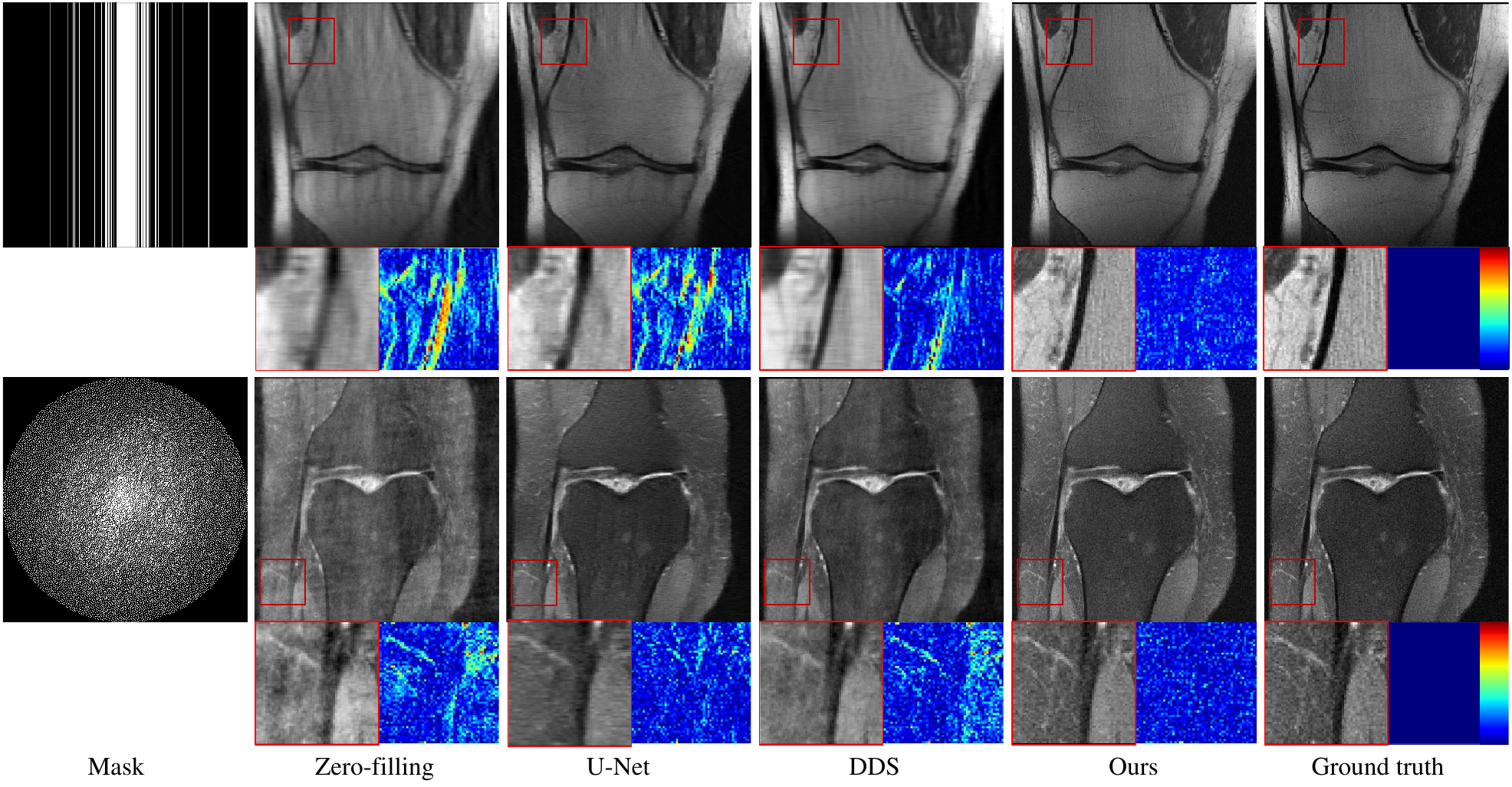}}
\caption{Qualitative results of fast MRI reconstruction.}
\label{fig:MRI}
\vspace{-4mm}
\end{figure*}

\begin{table}
\centering
\captionsetup{justification=centering, labelsep=newline} 
\renewcommand{\arraystretch}{1.7} % Increase row spacing
\caption{Quantitative results of sparse-view CT reconstruction.}
\begin{tabularx}{\linewidth}{>{\centering\arraybackslash}c|>{\centering\arraybackslash}X|>{\centering\arraybackslash}X|>{\centering\arraybackslash}X|>{\centering\arraybackslash}X|>{\centering\arraybackslash}X} % Center-align columns
\hline
                 & FBP & FBPConv & DOLCE & DDS & Ours \\ \hline\hline
\makecell{PSNR \\($\mathrm{dB}$)~$\uparrow$}
 & \makecell{16.5327 \\ (2.3140)} & \makecell{24.8138 \\ (2.5968)} & \makecell{18.3710 \\ (1.2488)} & \makecell{24.8110 \\ (3.0709)} & \makecell{\textbf{27.9169} \\ (4.1454)} \\ \hline
SSIM~$\uparrow$   & \makecell{0.6643 \\ (0.1352)} & \makecell{0.7756 \\ (0.1571)} & \makecell{0.7091 \\ (0.1396)} & \makecell{0.6488 \\ (0.1270)} & \makecell{\textbf{0.9297} \\ (0.0313)} \\ \hline
\end{tabularx}
\label{tab:SVCT}
% \vspace{-4mm}
\end{table}

\begin{table}[t]
\centering
\captionsetup{justification=centering, labelsep=newline} 
\renewcommand{\arraystretch}{1.7} % Increase row spacing
\caption{Quantitative results of MRI reconstruction.}
\begin{tabularx}{\linewidth}{>{\centering\arraybackslash}c|>{\centering\arraybackslash}X|>{\centering\arraybackslash}X|>{\centering\arraybackslash}X|>{\centering\arraybackslash}X} % Center-align columns
\hline
                 & Zero-filling & U-Net & DDS & Ours \\ \hline\hline
\makecell{PSNR \\($\mathrm{dB}$)~$\uparrow$}   & \makecell{29.4335 \\ (1.8688)} & \makecell{33.5105 \\ (2.2259)} & \makecell{30.4892 \\ (1.8758)} & \makecell{\textbf{34.5374} \\ (3.2673)}  \\ \hline
SSIM~$\uparrow$   & \makecell{0.8717\\ (0.0308)} & \makecell{0.9107 \\ (0.0453)} & \makecell{0.8847 \\ (0.0374)} & \makecell{\textbf{0.9130} \\ (0.0569)} \\ \hline
\end{tabularx}
\label{tab:MRI}
\vspace{-4mm}
\end{table}

\subsection{Downstream Tasks}
\label{sec:Downstream}
The pre-trained generative model can be adapted to a variety of downstream tasks through integration with ControlNet. 
In this section, we assess its performance across four tasks: sparse-view CT reconstruction, fast MRI reconstruction, data augmentation for segmentation and data augmentation for classification.

\subsubsection{Sparse-View CT Reconstruction}
\label{sec:Sparse-View CT Reconstruction}

In this experiment, we demonstrate the feasibility of leveraging the powerful generative prior knowledge learned by 3D MedDiffusion to address the sparse-view CT reconstruction.
We conduct experiment on a public dataset KiTs19~\cite{heller2019kits19}, which includes 210 Kidney CT scans.
For each 3D CT volume, we uniformly render 40 views in a full circle with fan-beam geometry based on ASTRA-toolbox~\cite{astratoolbox} and then perform reconstruction.
Our pre-trained model is fine-tuned with ControlNet.
The fine-tuning objective is defined in~\cref{eq:controlnet}. The Filtered Back-Projection (FBP)~\cite{fbp} reconstruction results used as the condition input with the full-view reconstruction results serving as the ground truth. 

We compare our method with several existing approaches, including FBPConv~\cite{jin2017deep}, DOLCE~\cite{liu2023dolce}, and DDS~\cite{chung2023decomposed}, and report both PSNR and SSIM metrics in~\cref{tab:SVCT} (with the values in parentheses denoting the standard deviations) to evaluate performance.
As shown in~\cref{tab:SVCT}, our method achieves the highest scores across both metrics, demonstrating our effectiveness.
Additionally, two representative cases are presented in Fig.~\ref{fig:SVCT}, where our reconstruction results deliver superior image quality and enhanced clarity of details.

\subsubsection{Fast MRI Reconstruction}

In this experiment, we aim to accelerate MRI reconstruction by leveraging pre-trained 3D MedDiffusion to map the k-space downsampled reconstructions to the fully-sampled reconstructions. 
We conduct experiment on \textbf{MRKnee} dataset and compare our method against zero-filling, U-Net~\cite{jin2017deep}, and DDS~\cite{chung2023decomposed}.
PSNR and SSIM are used as evaluation metrics. 
We adopt two types of downsampling masks including 1D Gaussian mask and Poisson mask, and the downsampling ratio is $8\times$.
The quantitative results are summarized in~\cref{tab:MRI} (with the values in parentheses denoting the standard deviations), with two examples of different downsampling masks shown in Fig.~\ref{fig:MRI}. 
Notably, our method achieves the highest metric performance in both downsampling masks, and obtain superior reconstructions.
Our volume-wise generation ensures better 3D consistency compared to the slice-wise generation of competing methods.

\begin{figure}[t]
\centerline{\includegraphics[width=\linewidth]{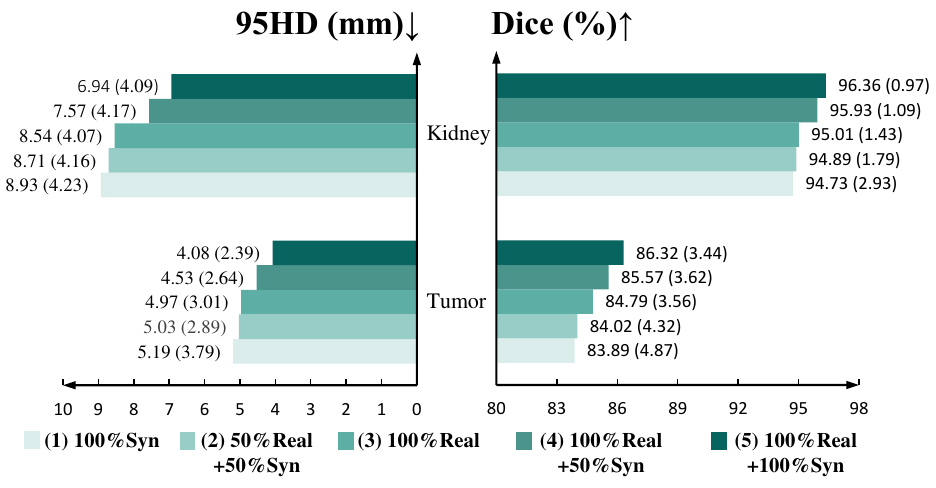}}
\caption{Quantitative comparison on data augmentation.}
\label{fig:data_aug}

\end{figure}

\begin{table}
\centering
\captionsetup{justification=centering, labelsep=newline} 
\renewcommand{\arraystretch}{1.4} 
\caption{Quantitative results of data augmentation for classification}
\begin{tabularx}{\linewidth}{>{\centering\arraybackslash}c|>{\centering\arraybackslash}X|>{\centering\arraybackslash}X}
\hline
                 & Baseline & Augmentation \\ \hline\hline
\makecell{Accuracy(\%)~$\uparrow$} & 75.00 & 79.17 \\ \hline
F1~$\uparrow$  & 0.7368 & 0.7863\\ \hline
\end{tabularx}
\label{tab:class}
\end{table}

\subsubsection{Data Augmentation for Segmentation}

Deep learning-based medical image segmentation is often limited by insufficient training data.
In this experiment, we leverage the strong generative capabilities of 3D MedDiffusion to augment the data and enhance segmentation performance.
We use a standard nnU-Net~\cite{isensee2018nnu} for segmentation on the KiTs19 dataset~\cite{heller2019kits19}, which includes kidney and tumor regions.
The testing set consists of 40 CT images and their segmentation masks, with the remaining 170 cases available for training.
We define the following five training settings:
\textbf{(1) 100\% Synthesis}: Generate new CT images using 3D MedDiffusion conditioned on the segmentation masks from the training set.
\textbf{(2) 50\% Real + 50\% Synthesis}: Randomly select 50\% of the real data from the training set, and then generate new CT images from the remaining 50\% using 3D MedDiffusion conditioned on their segmentation masks.
\textbf{(3) 100\% Real}: Use all 170 cases.
\textbf{(4) 100\% Real + 50\% Synthesis}: Use all 170 cases and generate new CT images for 50\% of the selected cases.
\textbf{(5) 100\% Real + 100\% Synthesis}: Use all 170 cases and generate new CT images for 100\% of the selected cases.

For evaluation, we conducted 5-fold cross-validation and used the 95\% Hausdorff Distance (95HD) and Dice Similarity Coefficient (DSC) as comparative metrics.
The overall quantitative results are presented in~\cref{fig:data_aug} (with the values in parentheses denoting the standard deviations).

For settings (1), (2), and (3), when the number of training samples is controlled, synthetic data can effectively replace real data.
Specifically, when comparing 100\% synthetic data with a mix of 50\% synthetic and 50\% real data (experiments (1) and (2)), the Dice score for tumor and kidney segmentation decreased by just 0.13\% and 0.16\%, respectively. Even when comparing 100\% synthetic data to 100\% real data (experiments (1) and (3)), the performance degradation remained minimal, with Dice score differences within 1\% and aslo 95HD variations within 1 mm for both tumor and kidney segmentation.

For settings (3), (4) and (5), even when the real data is relatively sufficient (100\%), the addition of synthetic data still improves the network's performance.
Specifically, when we introduce additional 100\% synthetic data, we observe a 1.6 mm decrease in 95HD for kidney segmentation and a 1.53\% increase in Dice score for tumor segmentation.

\begin{figure}[htbp]
\centering
\includegraphics[width=0.8\columnwidth]{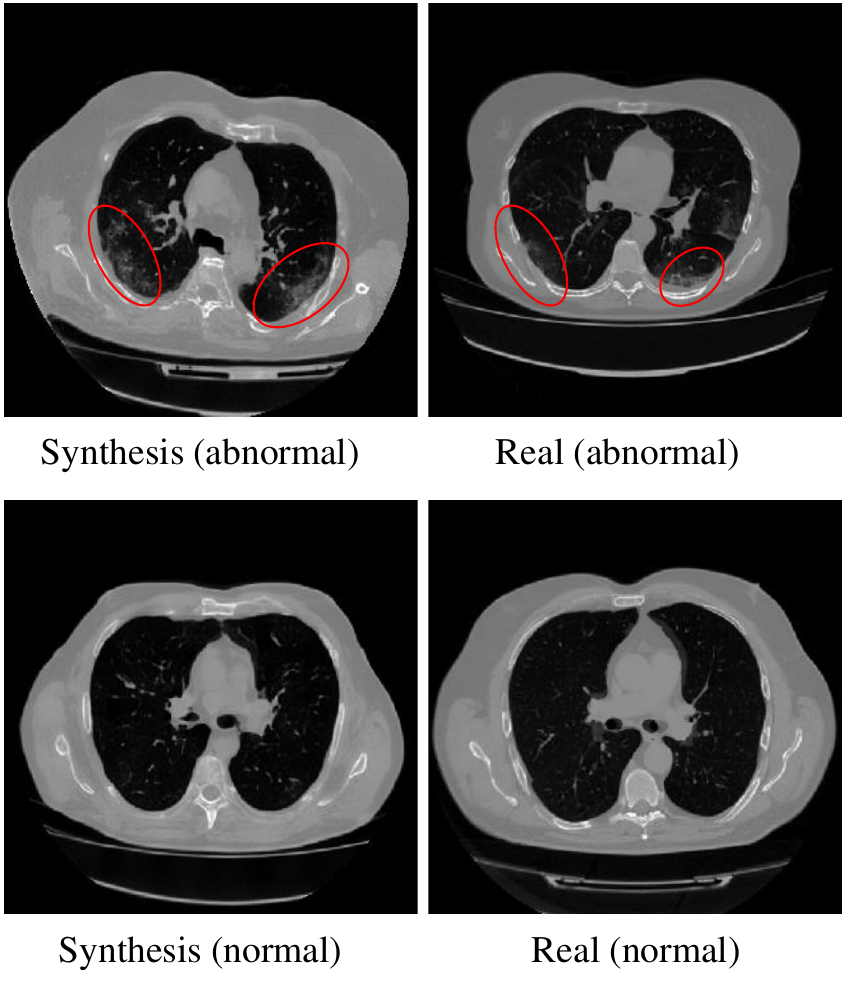}
\caption{Synthetic and real images, with (abnormal) and without (normal) COVID-19 related findings}
\label{figure:class_aug}
\end{figure}

\subsubsection{Data Augmentation for Classification}
In this experiment, we demonstrate the feasibility of leveraging the synthesized samples from generative model to augment the training dataset for a classification task. 
Different from the three downstream tasks above, in this task, we do not use the ControlNet but fine-tune our generative model on the MosMedData~\cite{morozov2020mosmeddata}, which consists of 1,110 studies, including both those with COVID-19 related findings and those without. The experiment was designed as follows: first, we fine-tuned our model on this dataset with minor modifications, using two class labels: $\mathbf{c} =0$ for studies with COVID-19 related findings (abnormal) and $\mathbf{c} =1$ for studies without COVID-19 findings (normal). Next, we used the trained model to generate new images with corresponding class labels (200 images for each of the two classes). Finally, we combined the original training dataset with the synthetic images to train a binary-class classifier as the augmented classifier, and used the original dataset without synthetic images to train another binary-class classifier as the baseline classifier. We then evaluated the performance of both classifiers on the test set. The dataset was split into training, validation, and test sets, with the test set containing containing 120 images. The model and classifiers were trained and evaluated on a single NVIDIA A100 GPU with 80GB of memory. We report the quantitative results in~\cref{tab:class}. Additionally, we show some synthetic images with and without COVID-19 related findings in~\cref{figure:class_aug}. It can be observed that our proposed method generates realistic normal and abnormal lung CT images. Moreover, the classifier trained with synthetic augmented data outperforms the baseline model, which was trained only on real images from the training set.

\section{Conclusion}
In this paper, we propose a 3D Medical Latent Diffusion (3D MedDiffusion) model to generate high-quality 3D medical images. 
3D MedDiffusion achieves its performance through two key design components: 1) the Patch-Volume Autoencoder and 2) the BiFlowNet noise estimator.
Additionally, it can be integrated with ControlNet, enabling efficient adaptation to a large range of downstream tasks in medical scenarios. 
Our experiments also demonstrate superior generative performance and strong generalizability across diverse tasks.

Although achieving promising results, our method still has limitations.
First, it lacks the ability to produce images at arbitrary resolutions.
Arbitrary-size generation could potentially be achieved with implicit neural encoder~\cite{kim2024arbitrary}, which will be our future work.
Second, our method does not incorporate age and gender as conditioning factors for synthesizing medical images. 
Future work could explore the inclusion of additional conditions to enhance clinical applicability.
Third, diffusion models require significant time and resources for generating high-resolution 3D medical images.
We aim to further improve its time and memory efficiency in the future.

{
\bibliographystyle{ieeetr}
\normalem
\bibliography{references}
}

\end{document}